\documentclass[12pt]{iopart}

\usepackage{iopams}  
\usepackage{bm}
\usepackage{here}
\usepackage{graphicx}
\usepackage{tikz}
\usepackage{xcolor}
\usepackage{circuitikz}
 \usepackage{pdflscape}
 \pgfcircdeclarebipolescaled{instruments}
{
    \anchor{text}{\pgfextracty{\pgf@circ@res@up}{\northeast}
        \pgfpoint{-.5\wd\pgfnodeparttextbox}{
            \dimexpr.5\dp\pgfnodeparttextbox+.5\ht\pgfnodeparttextbox+\pgf@circ@res@up\relax
        }
    }
}
{\ctikzvalof{bipoles/oscope/height}}
{josephson}
{\ctikzvalof{bipoles/oscope/height}}
{\ctikzvalof{bipoles/oscope/width}}
{
    \pgf@circ@setlinewidth{bipoles}{\pgfstartlinewidth}
    \pgfextracty{\pgf@circ@res@up}{\northeast}
    \pgfextractx{\pgf@circ@res@right}{\northeast}
    \pgfextractx{\pgf@circ@res@left}{\southwest}
    \pgfextracty{\pgf@circ@res@down}{\southwest}
    \pgfmathsetlength{\pgf@circ@res@step}{0.25*\pgf@circ@res@up}
    \pgfscope
        \pgfpathrectanglecorners{\pgfpoint{\pgf@circ@res@left}{\pgf@circ@res@down}}{\pgfpoint{\pgf@circ@res@right}{\pgf@circ@res@up}}
        \pgf@circ@draworfill
    \endpgfscope
    \pgfscope
      \pgfpathmoveto{\pgfpoint{\pgf@circ@res@left}{\pgf@circ@res@up}}%
      \pgfpathlineto{\pgfpoint{\pgf@circ@res@right}{\pgf@circ@res@down}}%
      \pgfpathmoveto{\pgfpoint{\pgf@circ@res@right}{\pgf@circ@res@up}}%
      \pgfpathlineto{\pgfpoint{\pgf@circ@res@left}{\pgf@circ@res@down}}%
      \pgfusepath{draw}
    \endpgfscope
}
\def\pgf@circ@josephson@path#1{\pgf@circ@bipole@path{josephson}{#1}}
\tikzset{josephson/.style = {\circuitikzbasekey, /tikz/to path=\pgf@circ@josephson@path, l=#1}}

\newcommand{\cev}[1]{\reflectbox{\ensuremath{\vec{\reflectbox{\ensuremath{#1}}}}}}

\begin{document}

\title[Constrained Hamiltonian dynamics for electrons]{Constrained Hamiltonian dynamics for electrons in magnetic field and additional forces besides the Lorentz force acting on electrons}
\author{Hiroyasu Koizumi}

\address{Division of Quantum Condensed Matter Physics, Center for Computational Sciences, University of Tsukuba, Tsukuba, Ibaraki 305-8577, Japan}
\ead{koizumi.hiroyasu.fn@u.tsukuba.ac.jp}
\vspace{10pt}
\begin{indented}
\item[]September, 2024
\end{indented}

\begin{abstract}
We consider the forces acting on electrons in magnetic field including the constraints and a condition arising from quantum mechanics. The force is calculated as the electron mass, $m_e$, multiplied
by the total time-derivative of the velocity field evaluated using the quantum mechanical many-electron wave function. The velocity field includes a term of the Berry connection from the many-body wave function; thereby, quantum mechanical effects are included.
It is shown that additional important forces besides the Lorentz force exist; they include the gradient of
the electron velocity field kinetic energy, the gradient of the chemical potential, and the `force' for producing topologically protected loop currents.
These additional forces are shown to be important in superconductivity,
electric current in metallic wires, and charging of capacitors.
\end{abstract}

%
%
%
%

\section{Introduction}

Currently, the basic equations for classical electromagnetism are
four Maxwell's equations and the Lorentz force acting on charged particles expressed using the electric field ${\bf E}$ and magnetic field ${\bf B}$ \cite{Lorentz1904,FeynmanII18-1}.
Although the above Maxwell's equations are named after Maxwell, the equations put forward by him were rather different \cite{Maxwell}; actually, he also derived the Lorentz force,
\begin{eqnarray}
{\bf F}_{\rm Lorentz}=q({\bf E}+{\bf v}\times{\bf B})
\label{eqLorentz}
\end{eqnarray}
where $q$ is the charge of the particle ($q=-e$ for electron)
 before Lorentz, in the course of obtaining the equation for the Faraday's law of induction,
 \begin{eqnarray}
{\cal E}=-{d \over {dt}}\int_{S(t)} {\bf B}({\bf r}, t)\cdot d{\bf S}
\label{Faraday}
\end{eqnarray}
where ${\cal E}$ is the electromotive force, $S(t)$ is the surface, ${\bf r}$ is the spatial coordinate, and $t$ is time \cite{Yaghjian2020}. The time dependence in the surface indicates the possible time-dependent movement of it.

He realized that the force expressed as a gradient of a scalar potential $\varphi$ was not enough to explain
the the Faraday's law of induction, and the vector potential ${\bf A}$ was introduced to go beyond the usual Newtonian description of the dynamics.
 With ${\bf A}$ and $\varphi$, ${\bf E}$ and ${\bf B}$ are given by
\begin{eqnarray}
{\bf E} &=& -\partial_t {\bf A}-\nabla \varphi
\nonumber
\\
{\bf B} &=& \nabla \times {\bf A}
\label{potentials}
\end{eqnarray}
Then, the electromotive force is expressed as
\begin{eqnarray}
{\cal E}=-{d \over {dt}}\int_{C(t)} {\bf A}({\bf r}, t)\cdot d{\bf r}
\label{Faraday2}
\end{eqnarray}
where $C(t)$ is the perimeter of $S(t)$.

Assuming the fluid dynamics of electric charge, the total time derivative of the vector potential 
\begin{eqnarray}
{d \over {dt}} {\bf A}({\bf r}, t)={\partial \over {\partial t}}{\bf A}({\bf r}, t)+{\bf v}({\bf r}, t) \cdot \nabla {\bf A}({\bf r}, t)
\label{eq1}
\end{eqnarray}
was considered with ${\bf v}({\bf r}, t)$ being the velocity field of the electric fluid.
Then, he showed that ${\bf E}$ arises from the first term, and ${\bf B}$ from the second.
Later, however, the vector potential was considered merely a mathematical tool, since the fluid that mediates electromagnetic force envisaged by
Maxwell (often called `ether') was not detected \cite{Lorentz1895}. Now all the forces acting on an electron is believed to be given by Eq.~(\ref{eqLorentz}) \cite{FeynmanII15-4}.

Next, let us consider an asymmetry in the electromotive force generation in a system composed of a magnet and a conductor in relative motion. One of the motivations that led Einstein to develop the special theory of relativity was this asymmetry
\cite{Einstein1905,Einstein1952}. The asymmetry is following: If the conductor is at rest and the magnet is moving, ${\bf E}$ is generated in the conductor; on the other hand, when the magnet is at rest and the conductor is moving, ${\bf B}$ is the cause of the electromotive force.
He showed that ${\bf E}$ and ${\bf B}$ appeared in Eq.~(\ref{eqLorentz}) were connected by the Lorentz transformation, thus, the asymmetry is just the difference in the frame used to describe the same phenomenon \cite{Einstein1905,Einstein1952}.
At this point, it was believed that standard four equations plus the Lorentz force expressed by ${\bf E}$ and ${\bf B}$ are the whole basic equations for electromagnetic phenomena \cite{FeynmanII18-1}.

However, Feynman claimed that the asymmetry still remains \cite{FeynmanII17-1}; if the conductor is at rest and the magnet is moving, ${\bf E}$  
calculated by the Maxwell-Faraday equation
\begin{eqnarray}
\nabla \times {\bf E} =- \partial_t {\bf B}
\label{Maxwell-Faraday}
\end{eqnarray}
is responsible for the electromotive force generation;
on the other hand, if the magnet is at rest and the conductor is moving, the
Lorentz force in Eq.~(\ref{eqLorentz}) with ${\bf B}$ is responsible for the electromotive force generation \cite{FeynmanII17-1}; thus, two different equations are used.
On the other hand, the Maxwell's derivation using ${\bf A}$ is a unified one, where two cases arise from 
the single ${\bf A}$; further, the Einstein's Lorentz transformation argument is already included in this view since the four vector $({ \varphi},{\bf A})$ satisfies the Lorentz transformation \cite{FeynmanII26-1}.

In quantum mechanics, ${\bf A}$ is a fundamental physical quantity describing electromagnetism as manifested in the Aharonov-Bohm effect \cite{AB1959,FeynmanII15-5} that was confirmed by experiments \cite{Tonomura1982,Tonomura1986}. Further, the velocity field of electrons exists as the expectation value of the velocity operator calculated by the many-body wave function. Thus, the quantum mechanical situation is more like the one Maxwell envisaged. 
Actually, the asymmetry pointed out by Feynman can be lifted by considering the duality of a $U(1)$ phase on the wave function that appears when electromagnetic field exists \cite{FluxRule}.

In quantum mechanics, an additional vector potential type object arises. They are called the `Berry connection ' \cite{Berry}. In particular, the 
Berry connection  from many-body wave functions,
\begin{eqnarray}
\!{\bf A}^{\rm MB}_{\Psi}({\bf r},t)\!=\!
{{{\rm Re} \left\{
 \int d\sigma_1  d{\bf x}_{2}  \cdots d{\bf x}_{N}
 \Psi^{\ast}({\bf r}, \sigma_1, \cdots, {\bf x}_{N},t)
  (-i \hbar \nabla )
\Psi({\bf r}, \sigma_1, \cdots, {\bf x}_{N},t) \right\}
 }
 \over {\hbar \rho({\bf r},t)}} 
 \nonumber
 \\
 \label{Berry}
\end{eqnarray}
defined by the present author is important for the electromagnetic problems. Originally, the Berry connection is defined for a wave function with adiabatic parameters. It can detect gauge structures hidden in the quantum system, particularly, it can detect defects in the parameter space that can modify the quantum behavior of the system in non-local (or holonomical) way. 
Even  lifting the adiabatic assumption of the parameter, it can detect the gauge structure and defects;
and here, we take the coordinate of a particle in many-body system as the parameter to calculate the Berry connection.
Thereby, the gauge structure and defects that arise in the coordinate space from many-body interactions can be detected.  
See Ref.~\cite{koizumi2020c} for detail of the derivation.
It has been argued to play crucial roles in superconductivity \cite{koizumi2022, koizumi2023}.
Here,
`$\rm{Re}$' denotes the real part, $\Psi$ is the total wave function, ${\bf x}_i$ collectively stands for the coordinate ${\bf r}_i$ and the spin $\sigma_i$ of the $i$th electron, $N$ is the total number of electrons, $-i \hbar \nabla$ is the Schr\"{o}dinger's momentum operator for the coordinate vector ${\bf r}$, and $\rho$ is the number density calculated from $\Psi$. 

In previous works \cite{koizumi2022, koizumi2023}, we have shown that the velocity field is given by
\begin{eqnarray}
{\bf v}={e \over m_e}{\bf A}+{\hbar \over {m_e}}{\bf A}_{\Psi}^{\rm MB}
\label{eq12v}
\end{eqnarray}
including ${\bf A}_{\Psi}^{\rm MB}$ \cite{Koizumi_2023}.
Then, the force acting on an electron is calculated by
\begin{eqnarray}
m_e {d \over {dt}}{\bf v}={e }{d \over {dt}}{\bf A}+{\hbar}{d \over {dt}}{\bf A}_{\Psi}^{\rm MB}
\label{eq3}
\end{eqnarray}
From the term $e{d \over {dt}}{\bf A}$, the Lorentz force arises. 

The main purpose of the present work is to show new forces arising from Eq.~(\ref{eq3}), and consider examples 
where they play important roles. In the due course, the meaning of the gauge invariance in the electromagnetism is changed;
in classical theory, all basic equations for electromagnetism are expressed using ${\bf E}$ and ${\bf B}$;
there exists a freedom in the choice (or gauge) of ${\bf A}$ and $\varphi$ for the same ${\bf E}$ and ${\bf B}$ in Eq.~(\ref{potentials}); thus, it is required that when a problem of electromagnetism is solved using ${\bf A}$ and $\varphi$, the result should not depend on the gauge used. 

The above gauge invariance requirement is handled differently in the present work. The velocity field in Eq.~(\ref{eq12v}) indicates that ${\bf A}_{\Psi}^{\rm MB}$ must compensate the gauge ambiguity in ${\bf A}$ and make the sum ${e \over m_e}{\bf A}+{\hbar \over {m_e}}{\bf A}_{\Psi}^{\rm MB}$ gauge invariant.
In this way, the gauge invariance requirement is now the requirement on ${\bf A}_{\Psi}^{\rm MB}$.
It is worth noting that a similar requirement arises in the Nambu-Goldstone mode appearing in the standard theory of
superconductivity \cite{Nambu1960}.
The requirement on ${\bf A}_{\Psi}^{\rm MB}$ can be achieved by imposing two constrains: 1) the conservation of local charge; 2) the single-valuedness of wave functions as functions of spatial coordinates. Actually this point was already explained in our previous work \cite{koizumi2023}. 
It is also important to know that ${\bf A}_{\Psi}^{\rm MB}$ gives rise to topologically protected loop currents, and supercurrent is generated by the velocity field in Eq.~(\ref{eq12v}) if those loop currents are stable \cite{Koizumi_2023}. This indicates ${\bf A}_{\Psi}^{\rm MB}$ replaces 
the  Nambu-Goldstone mode in the standard theory of superconductivity.

The constraints on ${\bf A}_{\Psi}^{\rm MB}$ are due to quantum mechanics. In other words, quantum mechanics requires additional constrains compared with classical mechanics. Then, additional forces arise from them. 
Such a possibility was already considered by Dirac in the theory of generalized Hamiltonian dynamics \cite{Dirac1958,Dirac2001}. He considered constrains expressed as functions of canonical conjugate variables;
however, the two constrains we have mentioned are more complicated ones.
We also add a condition in the generalized Hamiltonian dynamics; it is the presence of chemical potential (or Fermi energy). 
Metallic  electric current is generated by the chemical potential difference generated by an electric battery \cite{Landauer,Buttiker1985,Datta}, and the present formalism explains this current generation by including the term of the gradient of the chemical potential as a force arises in Eq.~(\ref{eq3}).

The organization of the present work is as follows: In Section~\ref{sec2}, two constraints and a condition arising from quantum mechanics are explained. In Section~\ref{sec3}, the Newton equation for the generalized Hamiltonian for electrons in electromagnetic field is presented and examined for some examples. Lastly, in Section~\ref{sec4}, we conclude the present work.

\section{Hamiltonian dynamics with two constraints and a condition from quantum mechanics}
\label{sec2}

We will consider two constrains and a condition on the generalized Hamiltonian dynamics required by quantum mechanics.
Some of the constrains are already considered in our previous work \cite{koizumi2022,koizumi2023}; however, we explain them succinctly, below, for completeness.

\subsection{Single-valuedness of the many-electron wave function as a function of electron coordinates}

In the Schr\"{o}dinger's representation, the operator for the momentum $\hat{p}_r$ conjugate to the canonical coordinate $\hat{q}_r$ is given by
\begin{eqnarray}
\hat{p}_r=-i \hbar {{\partial} \over {\partial q_r}}, \quad r=1, \cdots, n
\label{pr1}
\end{eqnarray}
This implicitly requires the existence of ket vectors 
\begin{eqnarray}
\hat{q}_r| q_1, \cdots, q_r, \cdots, q_n \rangle ={q}_r| q_1, \cdots, q_r, \cdots, q_n\rangle
\end{eqnarray}
that are eigenfunctions for the coordinate operators $\hat{q}_r$ \cite{DiracSec22}.
Since $q_r$ is an eigenvalue, it needs to be uniquely specified.  
The wave function for a physical state $| \psi \rangle$ is given by
\begin{eqnarray}
\psi(q_1, \cdots, q_r, \cdots, q_n) =\langle q_1, \cdots, q_r, \cdots, q_n | \psi \rangle
\end{eqnarray}
in the bra-ket notation.
Thus, the wave function is a single-valued function of $q_r$ due to the fact that the value of $q_r$ is uniquely specified.

The above single-valuedness constraint needs special attention when electrons perform spin-twisting itinerant motion. Since electrons are particle with spin ${1 \over 2}$, they may perform spin-twisting itinerant motion. Besides, their wave functions are spinors, sign-change may occur when the spin-twisting motion is such that it moves along a loop in the coordinate space with performing odd number of spin rotations.
In such a case, sign-change of wave functions occurs if the usual Schr\"{o}dinger equation is solved by the energy minimization requirement.
However, this sign-change is rectified if we introduce a $U(1)$ phase that compensates the sign change. Adding such a $U(1)$ phase is allowed in quantum mechanics, as is shown below; however, it is not well-known.
Usually, only cases with non sign-change are considered, thus, the $U(1)$ phase is a constant.
However, when dealing with cases with the sign-change, this $U(1)$ becomes an additional degree-of-freedom
needs to be handle.

Let us explain the appearance of the additional $U(1)$ phase, and how to handle when the sign-change occurs.
We start with the commutation relations in quantum mechanics given by
\begin{eqnarray}
[\hat{q}_r, \hat{q}_s]=0, \quad [\hat{p}_r, \hat{p}_s]=0, \quad  [\hat{q}_r, \hat{p}_s]=i \hbar \delta_{rs}
\end{eqnarray}
These relations are more fundamental than the Schr\"{o}dinger's representation in Eq.~(\ref{pr1}).
Thus, we may use the following $\hat{p}_r$'s 
\begin{eqnarray}
\hat{p}_r=-i \hbar {{\partial} \over {\partial q_r}}+{{\partial F} \over {\partial q_r}}, \quad r=1, \cdots, n
\label{pr2}
\end{eqnarray}
where $F$ is a function of $q_1, \cdots, q_n$, and also of time $t$. 

Dirac claimed that this extra ${{\partial F} \over {\partial q_r}}$ is, however, can be removed by
the following transformation of the wave function,
\begin{eqnarray}
\psi(q_1, \cdots, q_n) \rightarrow e^{i \gamma(q_1, \cdots, q_n, t)} \psi(q_1, \cdots, q_n) 
\label{gamma}
\end{eqnarray}
with $\gamma$ related to $F$ by
$
F(q_1, \cdots, q_n, t)=-\hbar \gamma(q_1, \cdots, q_n, t) + \mbox{constant}
$.
Thus, we can always use $p_r$'s in Eq.~(\ref{pr1}) and deal with the wave function $e^{i \gamma} \psi(q_1, \cdots, q_n)$ \cite{DiracSec22}. The wave function $e^{i \gamma} \psi(q_1, \cdots, q_n)$ is taken up as a whole in the usual treatment, thus, $e^{i \gamma}$ is not considered.

Contrary to the Dirac's claim, there exist situations where this $e^{i \gamma}$ as an
extra degree-of-freedom that can be used to fulfill the single-valuedness constraint \cite{koizumi2022}. 
Note that this type of $U(1)$ phase was first introduced by O'Brien in the context of the dynamical Jahn-Teller problem, where the degenerate electronic wave function by symmetry behaves as a spinor \cite{OBrien1964}.

The relation $[p_r, p_s]=0$ in Eq.~(\ref{pr2}) requires the following condition
\begin{eqnarray}
{\partial \over {\partial q_s}}{{\partial F} \over {\partial q_r}}-{\partial \over {\partial q_r}}{{\partial F} \over {\partial q_s}}=0
\label{eqSingular}
\end{eqnarray}
In the case where $e^{i \gamma}$ is multi-valued, this condition is violated at some points. They are topological defects where the amplitude of $\psi(q_1, \cdots, q_n)$ is zero.
Due to this zeroness of the amplitude, we can treat them as singularities. Such defects arise when electrons perform spin-twisting itinerant motion with centers of the spin-twisting being the defects.

We rewrite  $e^{i \gamma} \psi(q_1, \cdots, q_n)$ in Eq.~(\ref{gamma}) as
\begin{eqnarray}
\Psi ({\bf x}_1, \cdots, {\bf x}_{N},t)=\Psi_0 ({\bf x}_1, \cdots, {\bf x}_{N},t)\exp\left(i \sum_{j=1}^{N} \int_{0}^{{\bf r}_j} {\bf A}_{\Psi}^{\rm MB}({\bf r}',t) \cdot d{\bf r}' \right)
\nonumber
\\
\label{wavef0}
\end{eqnarray}
so that the relation to the Berry connection in Eq.~(\ref{Berry}) becomes apparent;
we can identify $\gamma$ as
\begin{eqnarray}
\gamma({\bf r}_1, \cdots, {\bf r}_N, t)= \sum_{j=1}^{N} \int_0^{{\bf r}_j} {\bf A}^{\rm MB}_{\Psi}({\bf r}', t) \cdot d{\bf r}' 
\label{gamma1}
\end{eqnarray}
For convenience sake, we introduce angular variable $\chi$ with period $2\pi$,
 \begin{eqnarray}
{ {\chi({\bf r},t)}}= - 2\int^{{\bf r}}_0 {\bf A}_{\Psi}^{\rm MB}({\bf r}',t) \cdot d{\bf r}' 
\end{eqnarray}
The multi-valued behavior of $\chi$ is characterized by the topological integer, `winding number', defined by
\begin{eqnarray}
w_C[\chi]={1 \over {2\pi}}\oint_C \nabla \chi \cdot d{\bf r}
\label{winding}
\end{eqnarray}
The winding number $w_C[\chi]$ becomes non-zero integer when singularities of $\chi$ exist within the loop $C$.
The wave function $\Psi$ is now expressed as
\begin{eqnarray}
\Psi ({\bf x}_1, \cdots, {\bf x}_{N},t)=\Psi_0 ({\bf x}_1, \cdots, {\bf x}_{N},t)\exp\left(-{i \over 2}\sum_{j=1}^{N}\chi({\bf r}_j,t)\right)
\nonumber
\\
\label{wavef1}
\end{eqnarray}
 The velocity field ${\bf v}$ is calculated as the expectation value of the velocity field operator (see Appendix \ref{velocityfield} for definition),
\begin{eqnarray}
{\bf v}={e \over m_e}{\bf A}+{\hbar \over {m_e}}{\bf A}_{\Psi}^{\rm MB}
={e \over m_e}\left({\bf A}-{\hbar \over {2e}}\nabla \chi \right)
\label{eq12v2}
\end{eqnarray}

\begin{figure}[H]
  \begin{center}
  \newcommand{\Depth}{2}
\newcommand{\Height}{2}
\newcommand{\Width}{2}
  \begin{tikzpicture}
\coordinate (X000) at (0,0,0);
\coordinate (X001) at (0,0,\Height);
\coordinate (X002) at (0,0,2*\Height);
\coordinate (X003) at (0,0,3*\Height);

\coordinate (X010) at (0,\Width,0);
\coordinate (X011) at (0,\Width,\Height);
\coordinate (X012) at (0,\Width,2*\Height);
\coordinate (X013) at (0,\Width,3*\Height);

\coordinate (X020) at (0,2*\Width,0);
\coordinate (X021) at (0,2*\Width,\Height);
\coordinate (X022) at (0,2*\Width,2*\Height);
\coordinate (X023) at (0,2*\Width,3*\Height);

\coordinate (X030) at (0,3*\Width,0);
\coordinate (X031) at (0,3*\Width,\Height);
\coordinate (X032) at (0,3*\Width,2*\Height);
\coordinate (X033) at (0,3*\Width,3*\Height);

\coordinate (X100) at (\Depth,0,0);
\coordinate (X101) at (\Depth,0,\Height);
\coordinate (X102) at (\Depth,0,2*\Height);
\coordinate (X103) at (\Depth,0,3*\Height);

\coordinate (X110) at (\Depth,\Width,0);
\coordinate (X111) at (\Depth,\Width,\Height);
\coordinate (X112) at (\Depth,\Width,2*\Height);
\coordinate (X113) at (\Depth,\Width,3*\Height);

\coordinate (X120) at (\Depth,2*\Width,0);
\coordinate (X121) at (\Depth,2*\Width,\Height);
\coordinate (X122) at (\Depth,2*\Width,2*\Height);
\coordinate (X123) at (\Depth,2*\Width,3*\Height);

\coordinate (X130) at (\Depth,3*\Width,0);
\coordinate (X131) at (\Depth,3*\Width,\Height);
\coordinate (X132) at (\Depth,3*\Width,2*\Height);
\coordinate (X133) at (\Depth,3*\Width,3*\Height);

\coordinate (X200) at (2*\Depth,0,0);
\coordinate (X201) at (2*\Depth,0,\Height);
\coordinate (X202) at (2*\Depth,0,2*\Height);
\coordinate (X203) at (2*\Depth,0,3*\Height);

\coordinate (X210) at (2*\Depth,\Width,0);
\coordinate (X211) at (2*\Depth,\Width,\Height);
\coordinate (X212) at (2*\Depth,\Width,2*\Height);
\coordinate (X213) at (2*\Depth,\Width,3*\Height);

\coordinate (X220) at (2*\Depth,2*\Width,0);
\coordinate (X221) at (2*\Depth,2*\Width,\Height);
\coordinate (X222) at (2*\Depth,2*\Width,2*\Height);
\coordinate (X223) at (2*\Depth,2*\Width,3*\Height);

\coordinate (X230) at (2*\Depth,3*\Width,0);
\coordinate (X231) at (2*\Depth,3*\Width,\Height);
\coordinate (X232) at (2*\Depth,3*\Width,2*\Height);
\coordinate (X233) at (2*\Depth,3*\Width,3*\Height);

\coordinate (X300) at (3*\Depth,0,0);
\coordinate (X301) at (3*\Depth,0,\Height);
\coordinate (X302) at (3*\Depth,0,2*\Height);
\coordinate (X303) at (3*\Depth,0,3*\Height);

\coordinate (X310) at (3*\Depth,\Width,0);
\coordinate (X311) at (3*\Depth,\Width,\Height);
\coordinate (X312) at (3*\Depth,\Width,2*\Height);
\coordinate (X313) at (3*\Depth,\Width,3*\Height);

\coordinate (X320) at (3*\Depth,2*\Width,0);
\coordinate (X321) at (3*\Depth,2*\Width,\Height);
\coordinate (X322) at (3*\Depth,2*\Width,2*\Height);
\coordinate (X323) at (3*\Depth,2*\Width,3*\Height);

\coordinate (X330) at (3*\Depth,3*\Width,0);
\coordinate (X331) at (3*\Depth,3*\Width,\Height);
\coordinate (X332) at (3*\Depth,3*\Width,2*\Height);
\coordinate (X333) at (3*\Depth,3*\Width,3*\Height);

\draw[fill=red!5] (X000) -- (X001) -- (X101) -- (X100) -- cycle;
\draw[fill=red!5] (X001) -- (X002) -- (X102) -- (X101) -- cycle;
\draw[fill=red!5] (X002) -- (X003) -- (X103) -- (X102) -- cycle;
\draw[fill=red!5] (X000) -- (X001) -- (X201) -- (X200) -- cycle;
\draw[fill=red!5] (X001) -- (X002) -- (X202) -- (X201) -- cycle;
\draw[fill=red!5] (X002) -- (X003) -- (X203) -- (X202) -- cycle;
\draw[fill=red!5] (X200) -- (X201) -- (X301) -- (X300) -- cycle;
\draw[fill=red!5] (X201) -- (X202) -- (X302) -- (X301) -- cycle;
\draw[fill=red!5] (X202) -- (X203) -- (X303) -- (X302) -- cycle;

\draw[fill=red!5] (X000) -- (X010) -- (X110) -- (X100) -- cycle;
\draw[fill=red!5] (X010) -- (X020) -- (X120) -- (X110) -- cycle;
\draw[fill=red!5] (X020) -- (X030) -- (X130) -- (X120) -- cycle;
\draw[fill=red!5] (X200) -- (X210) -- (X110) -- (X100) -- cycle;
\draw[fill=red!5] (X210) -- (X220) -- (X120) -- (X110) -- cycle;
\draw[fill=red!5] (X220) -- (X230) -- (X130) -- (X120) -- cycle;
\draw[fill=red!5] (X300) -- (X310) -- (X210) -- (X200) -- cycle;
\draw[fill=red!5] (X310) -- (X320) -- (X220) -- (X210) -- cycle;
\draw[fill=red!5] (X320) -- (X330) -- (X230) -- (X220) -- cycle;

\draw[fill=red!5] (X000) -- (X010) -- (X011) -- (X001) -- cycle;
\draw[fill=red!5] (X010) -- (X020) -- (X021) -- (X011) -- cycle;
\draw[fill=red!5] (X020) -- (X030) -- (X031) -- (X021) -- cycle;
\draw[fill=red!5] (X001) -- (X011) -- (X012) -- (X002) -- cycle;
\draw[fill=red!5] (X011) -- (X021) -- (X022) -- (X012) -- cycle;
\draw[fill=red!5] (X021) -- (X031) -- (X032) -- (X022) -- cycle;
\draw[fill=red!5] (X002) -- (X012) -- (X013) -- (X003) -- cycle;
\draw[fill=red!5] (X012) -- (X022) -- (X023) -- (X013) -- cycle;
\draw[fill=red!5] (X022) -- (X032) -- (X033) -- (X023) -- cycle;

\draw[fill=red!5,opacity=0.6] (X010) -- (X011) -- (X111) -- (X110) -- cycle;
\draw[fill=red!5,opacity=0.6] (X011) -- (X012) -- (X112) -- (X111) -- cycle;
\draw[fill=red!5,opacity=0.6] (X012) -- (X013) -- (X113) -- (X112) -- cycle;
\draw[fill=red!5,opacity=0.6] (X010) -- (X011) -- (X211) -- (X210) -- cycle;
\draw[fill=red!5,opacity=0.6] (X011) -- (X012) -- (X212) -- (X211) -- cycle;
\draw[fill=red!5,opacity=0.6] (X012) -- (X013) -- (X213) -- (X212) -- cycle;
\draw[fill=red!5,opacity=0.6] (X210) -- (X211) -- (X311) -- (X310) -- cycle;
\draw[fill=red!5,opacity=0.6] (X211) -- (X212) -- (X312) -- (X311) -- cycle;
\draw[fill=red!5,opacity=0.6] (X212) -- (X213) -- (X313) -- (X312) -- cycle;

\draw[fill=red!5,opacity=0.6] (X020) -- (X021) -- (X121) -- (X120) -- cycle;
\draw[fill=red!5,opacity=0.6] (X021) -- (X022) -- (X122) -- (X121) -- cycle;
\draw[fill=red!5,opacity=0.6] (X022) -- (X023) -- (X123) -- (X122) -- cycle;
\draw[fill=red!5,opacity=0.6] (X020) -- (X021) -- (X221) -- (X220) -- cycle;
\draw[fill=red!5,opacity=0.6] (X021) -- (X022) -- (X222) -- (X221) -- cycle;
\draw[fill=red!5,opacity=0.6] (X022) -- (X023) -- (X223) -- (X222) -- cycle;
\draw[fill=red!5,opacity=0.6] (X220) -- (X221) -- (X321) -- (X320) -- cycle;
\draw[fill=red!5,opacity=0.6] (X221) -- (X222) -- (X322) -- (X321) -- cycle;
\draw[fill=red!5,opacity=0.6] (X222) -- (X223) -- (X323) -- (X322) -- cycle;

\draw[fill=red!5,opacity=0.6] (X001) -- (X011) -- (X111) -- (X101) -- cycle;
\draw[fill=red!5,opacity=0.6] (X011) -- (X021) -- (X121) -- (X111) -- cycle;
\draw[fill=red!5,opacity=0.6] (X021) -- (X031) -- (X131) -- (X121) -- cycle;
\draw[fill=red!5,opacity=0.6] (X101) -- (X111) -- (X211) -- (X201) -- cycle;
\draw[fill=red!5,opacity=0.6] (X111) -- (X121) -- (X221) -- (X211) -- cycle;
\draw[fill=red!5,opacity=0.6] (X121) -- (X131) -- (X231) -- (X221) -- cycle;
\draw[fill=red!5,opacity=0.6] (X201) -- (X211) -- (X311) -- (X301) -- cycle;
\draw[fill=red!5,opacity=0.6] (X211) -- (X221) -- (X321) -- (X311) -- cycle;
\draw[fill=red!5,opacity=0.6] (X221) -- (X231) -- (X331) -- (X321) -- cycle;

\draw[fill=red!5,opacity=0.6] (X002) -- (X012) -- (X112) -- (X102) -- cycle;
\draw[fill=red!5,opacity=0.6] (X012) -- (X022) -- (X122) -- (X112) -- cycle;
\draw[fill=red!5,opacity=0.6] (X022) -- (X032) -- (X132) -- (X122) -- cycle;
\draw[fill=red!5,opacity=0.6] (X102) -- (X112) -- (X212) -- (X202) -- cycle;
\draw[fill=red!5,opacity=0.6] (X112) -- (X122) -- (X222) -- (X212) -- cycle;
\draw[fill=red!5,opacity=0.6] (X122) -- (X132) -- (X232) -- (X222) -- cycle;
\draw[fill=red!5,opacity=0.6] (X202) -- (X212) -- (X312) -- (X302) -- cycle;
\draw[fill=red!5,opacity=0.6] (X212) -- (X222) -- (X322) -- (X312) -- cycle;
\draw[fill=red!5,opacity=0.6] (X222) -- (X232) -- (X332) -- (X322) -- cycle;

\draw[fill=black!50,opacity=0.6] (X100) -- (X110) -- (X111) -- (X101) -- cycle;
\draw[fill=black!50,opacity=0.6] (X110) -- (X120) -- (X121) -- (X111) -- cycle;
\draw[fill=black!50,opacity=0.6] (X120) -- (X130) -- (X131) -- (X121) -- cycle;
\draw[fill=black!50,opacity=0.6] (X101) -- (X111) -- (X112) -- (X102) -- cycle;
\draw[fill=black!50,opacity=0.6] (X111) -- (X121) -- (X122) -- (X112) -- cycle;
\draw[fill=black!50,opacity=0.6] (X121) -- (X131) -- (X132) -- (X122) -- cycle;
\draw[fill=black!50,opacity=0.6] (X102) -- (X112) -- (X113) -- (X103) -- cycle;
\draw[fill=black!50,opacity=0.6] (X112) -- (X122) -- (X123) -- (X113) -- cycle;
\draw[fill=black!50,opacity=0.6] (X122) -- (X132) -- (X133) -- (X123) -- cycle;

\draw[fill=black!50,opacity=0.6] (X200) -- (X210) -- (X211) -- (X201) -- cycle;
\draw[fill=black!50,opacity=0.6] (X210) -- (X220) -- (X221) -- (X211) -- cycle;
\draw[fill=black!50,opacity=0.6] (X220) -- (X230) -- (X231) -- (X221) -- cycle;
\draw[fill=black!50,opacity=0.6] (X201) -- (X211) -- (X212) -- (X202) -- cycle;
\draw[fill=black!50,opacity=0.6] (X211) -- (X221) -- (X222) -- (X212) -- cycle;
\draw[fill=black!50,opacity=0.6] (X221) -- (X231) -- (X232) -- (X222) -- cycle;
\draw[fill=black!50,opacity=0.6] (X202) -- (X212) -- (X213) -- (X203) -- cycle;
\draw[fill=black!50,opacity=0.6] (X212) -- (X222) -- (X223) -- (X213) -- cycle;
\draw[fill=black!50,opacity=0.6] (X222) -- (X232) -- (X233) -- (X223) -- cycle;

\draw[fill=red!5,opacity=0.6] (X003) -- (X013) -- (X113) -- (X103) -- cycle;
\draw[fill=red!5,opacity=0.6] (X013) -- (X023) -- (X123) -- (X113) -- cycle;
\draw[fill=red!5,opacity=0.6] (X023) -- (X033) -- (X133) -- (X123) -- cycle;
\draw[fill=red!5,opacity=0.6] (X103) -- (X113) -- (X213) -- (X203) -- cycle;
\draw[fill=red!5,opacity=0.6] (X113) -- (X123) -- (X223) -- (X213) -- cycle;
\draw[fill=red!5,opacity=0.6] (X123) -- (X133) -- (X233) -- (X223) -- cycle;
\draw[fill=red!5,opacity=0.6] (X203) -- (X213) -- (X313) -- (X303) -- cycle;
\draw[fill=red!5,opacity=0.6] (X213) -- (X223) -- (X323) -- (X313) -- cycle;
\draw[fill=red!5,opacity=0.6] (X223) -- (X233) -- (X333) -- (X323) -- cycle;

\draw[fill=black!50,opacity=0.6] (X300) -- (X310) -- (X311) -- (X301) -- cycle;
\draw[fill=black!50,opacity=0.6] (X310) -- (X320) -- (X321) -- (X311) -- cycle;
\draw[fill=black!50,opacity=0.6] (X320) -- (X330) -- (X331) -- (X321) -- cycle;
\draw[fill=black!50,opacity=0.6] (X301) -- (X311) -- (X312) -- (X302) -- cycle;
\draw[fill=black!50,opacity=0.6] (X311) -- (X321) -- (X322) -- (X312) -- cycle;
\draw[fill=black!50,opacity=0.6] (X321) -- (X331) -- (X332) -- (X322) -- cycle;
\draw[fill=black!50,opacity=0.6] (X302) -- (X312) -- (X313) -- (X303) -- cycle;
\draw[fill=black!50,opacity=0.6] (X312) -- (X322) -- (X323) -- (X313) -- cycle;
\draw[fill=black!50,opacity=0.6] (X322) -- (X332) -- (X333) -- (X323) -- cycle;

\draw[fill=red!5,opacity=0.6] (X030) -- (X031) -- (X131) -- (X130) -- cycle;
\draw[fill=red!5,opacity=0.6] (X031) -- (X032) -- (X132) -- (X131) -- cycle;
\draw[fill=red!5,opacity=0.6] (X032) -- (X033) -- (X133) -- (X132) -- cycle;
\draw[fill=red!5,opacity=0.6] (X030) -- (X031) -- (X231) -- (X230) -- cycle;
\draw[fill=red!5,opacity=0.6] (X031) -- (X032) -- (X232) -- (X231) -- cycle;
\draw[fill=red!5,opacity=0.6] (X032) -- (X033) -- (X233) -- (X232) -- cycle;
\draw[fill=red!5,opacity=0.6] (X230) -- (X231) -- (X331) -- (X330) -- cycle;
\draw[fill=red!5,opacity=0.6] (X231) -- (X232) -- (X332) -- (X331) -- cycle;
\draw[fill=red!5,opacity=0.6] (X232) -- (X233) -- (X333) -- (X332) -- cycle;

\end{tikzpicture}
  \end{center}
    \caption{Another example of flattening of a 3D lattice. The eight shaded faces in the 3D lattice are removed. $N_b=54, N_p=28, N_v=27$ for this case.}
  \label{3D2D2}
\end{figure}
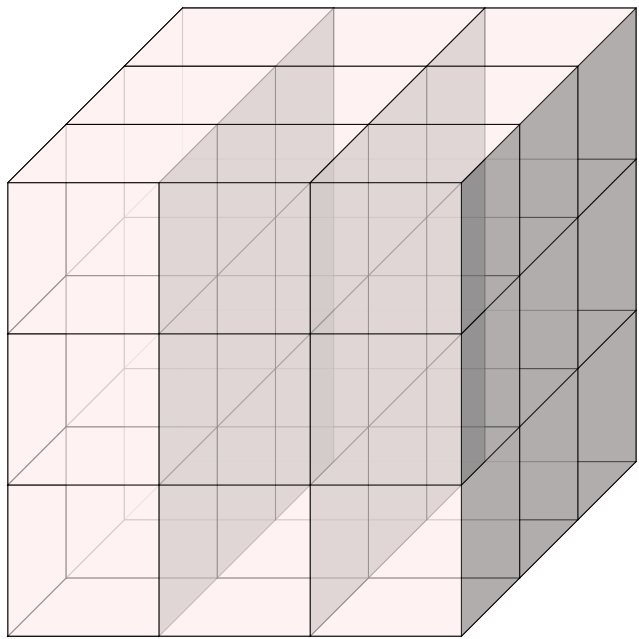

Now we shall consider the single-valuedness constraint, explicitly. It is taken as
the constraint on winding numbers of $\chi$ around loops.
The loops for calculating them are constructed in the following way:
Let us consider `flattening' of a 3D lattice; examples are depicted in Figs.~\ref{3D2D1} and \ref{3D2D2}.
The `flattening' is a procedure in which some faces of the 3D lattice are removed so that 
 there is no cell surrounded by faces after the removing.
 For the flattened lattice (i.e., the lattice obtained by flattening),
there exists a path from outside to the center of every cell.
For the flattened lattice, the following equality holds,
\begin{eqnarray}
N_b = N_p + N_v -1
\label{eqNb}
\end{eqnarray}
where $N_b$ is the total number of bonds (or edges), $N_p$ is the total number of plaques (or faces), and
$N_v$ is the total number of vertices (or sites).
The above equality is the Euler's relation for a 2D lattice; thus,
we call the above procedure as the `flattening'. The same flattening is considered in
our previous work, where the flattened lattices are 2D lattices in the 2D plane \cite{koizumi2023}.

\begin{figure}[H]
  \begin{center}
  \newcommand{\Depth}{2}
\newcommand{\Height}{2}
\newcommand{\Width}{2}
  \begin{tikzpicture}
\coordinate (X000) at (0,0,0);

\coordinate (X010) at (0,\Width,0);
\coordinate (X020) at (0,2*\Width,0);

\coordinate (X011) at (0,\Width,\Height);
\coordinate (X021) at (0,2*\Width,\Height);

\coordinate (X001) at (0,0,\Height);
\coordinate (X100) at (\Depth,0,0);

\coordinate (X110) at (\Depth,\Width,0);
\coordinate (X120) at (\Depth,2*\Width,0);

\coordinate (X111) at (\Depth,\Width,\Height);
\coordinate (X121) at (\Depth,2*\Width,\Height);

\coordinate (X101) at (\Depth,0,\Height);
\coordinate (X200) at (2*\Depth,0,0);

\coordinate (X210) at (2*\Depth,\Width,0);
\coordinate (X220) at (2*\Depth,2*\Width,0);

\coordinate (X211) at (2*\Depth,\Width,\Height);
\coordinate (X221) at (2*\Depth,2*\Width,\Height);

\coordinate (X201) at (2*\Depth,0,\Height);

\draw[fill=red!5] (X000) -- (X001) -- (X101) -- (X100) -- cycle;
\draw[fill=red!5] (X100) -- (X101) -- (X201) -- (X200) -- cycle;

\draw[fill=red!5,opacity=0.6] (X010) -- (X011) -- (X111) -- (X110) -- cycle;

\draw[fill=red!5,opacity=0.6] (X010) -- (X011) -- (X211) -- (X210) -- cycle;

\draw[fill=red!5] (X000) -- (X010) -- (X110) -- (X100) -- cycle;
\draw[fill=red!5] (X010) -- (X020) -- (X120) -- (X110) -- cycle;

\draw[fill=red!5] (X200) -- (X210) -- (X110) -- (X100) -- cycle;
\draw[fill=red!5] (X210) -- (X220) -- (X120) -- (X110) -- cycle;

\draw[fill=red!5] (X000) -- (X010) -- (X011) -- (X001) -- cycle;
\draw[fill=red!5] (X010) -- (X020) -- (X021) -- (X011) -- cycle;


\draw[fill=red!5,opacity=0.6] (X001) -- (X011) -- (X111) -- (X101) -- cycle;
\draw[fill=red!5,opacity=0.6] (X011) -- (X021) -- (X121) -- (X111) -- cycle;

\draw[fill=black!50,opacity=0.6] (X200) -- (X210) -- (X211) -- (X201) -- cycle;
\draw[fill=black!50,opacity=0.6] (X210) -- (X220) -- (X221) -- (X211) -- cycle;

\draw[fill=red!5,opacity=0.6] (X001) -- (X011) -- (X211) -- (X201) -- cycle;
\draw[fill=red!5,opacity=0.6] (X011) -- (X021) -- (X221) -- (X211) -- cycle;

\draw[fill=red!5,opacity=0.6] (X020) -- (X021) -- (X121) -- (X120) -- cycle;
\draw[fill=red!5,opacity=0.6] (X020) -- (X021) -- (X221) -- (X220) -- cycle;
\end{tikzpicture}
  \end{center}
    \caption{An example for flattening a 3D lattice. The four shaded faces in the 3D lattice are removed. $N_b=33, N_p=16, N_v=18$ for this case.}
  \label{3D2D1}
\end{figure}
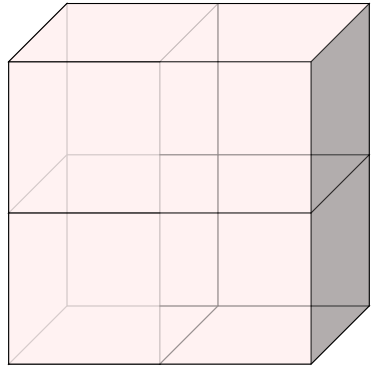

The singularities of $\chi$ form strings.
Analogous string of singularities is considered by Dirac \cite{Monopole}.
The Dirac's string has terminal points, but, the strings here are those without terminal points. 
When a closed path $C$ is
taken around a string, we have the winding number given in Eq.~(\ref{winding}).
If this number is odd, $e^{-{i \over 2}\chi}$ (the $U(1)$ phase factor appearing in Eq.~(\ref{wavef1})) changes sign after the shifting of the coordinate once around $C$.
We assume at most one string enters each cell from a face of the cell and exits from another face of the same cell.
The winding numbers of $\chi$ for all the plaques of the flattened lattice are given as externally supplied parameters. In other words, they are boundary conditions. 
They are chosen so that the single-valuedness of the total wave function is achieved.
From the single-valuedness constraint, totally $N_p$ equations are obtained.

\subsection{Conservation of local charge constraint}

In order to specify $\nabla \chi$ in the lattice, we need to know $N_b$ values of the difference of $\chi$ values at sites across each bond. 
The relation in Eq.~(\ref{eqNb}) indicates that we need $N_v-1$ additional equations besides $N_p$ equations from the single-valuedness constraint.
Actually, the conservation of local charge constraint supplies
the required number of equations: The number of points where the conservation is considered is $N_v$, however, the total charge conservation is imposed, separately; thus,
$N_v$ equations are redundant, and $N_v-1$ equations are independent, where $-1$ arises from the total charge conservation.
With the two types of constrains, one from the single-valuedness of the wave function and the other from the conservation of the charge, the discrete version of $\nabla \chi$ are uniquely obtained.
 
Let us calculate the total time-derivative of the velocity field ${\bf v}$,
\begin{eqnarray}
{{d{\bf v}} \over {dt}}&=&\partial_t{\bf v}+({\bf v} \cdot \nabla){\bf v}
\nonumber
\\
&=&\partial_t{\bf v}+
{1 \over 2}\nabla v^2-{\bf v}\times(\nabla \times {\bf v})
\nonumber
\\
&=&{ e \over {m_e}} \partial_t {\bf A}-{ \hbar \over {2m_e}}\partial_t(\nabla \chi)
+{1 \over 2}\nabla v^2-{ e \over m_e}{\bf v}\times{\bf B}
+{ \hbar \over {2m_e}}{\bf v}\times \nabla \times (\nabla \chi)
\nonumber
\\
\label{eq13a}
\end{eqnarray}
where ${\bf B}=\nabla \times {\bf A}$ is use. The obtained $\nabla \chi$ can be used in the discretized version of the above formula.

\subsection{The existence of chemical potential}

In order to deal with electric current conduction in metallic wires, we include the chemical potential.
The presence of the chemical potential is a quantum effect arising from the existence of the Fermi energy in metals.

First, we use the relation ${\bf E}=-\partial_t {\bf A} -\nabla \varphi$ in
Eq.~(\ref{eq13a}),
\begin{eqnarray}
{{d{\bf v}} \over {dt}}
&=&{ {-e}\over {m_e}}{\bf E}+{ {-e} \over {m_e}}\nabla \left(
\varphi + { \hbar \over {2e}}\partial_t \chi
\right)
+{1 \over 2}\nabla v^2-{ e \over m_e}{\bf v}\times{\bf B}
\nonumber
\\
&&
+{ \hbar \over {2m_e}}{\bf v}\times \nabla \times (\nabla \chi)
+{ \hbar \over {2m_e}}(\nabla \partial_t -\partial_t \nabla )\chi
\nonumber
\\
\label{eq24}
\end{eqnarray}
Next, we consider the sum, ${\bf A}-{\hbar \over {2e}}\nabla \chi$, in ${\bf v}$.
This sum is gauge invariant, which also implies its time-component partner
${\varphi}+{\hbar \over {2e}}\partial_t \chi $
is gauge invariant. 
Taking into account the fact that the chemical potential $\mu$ appears in a similar manner as the 
scalar potential $\varphi$ in the electron Hamiltonian, and also the fact that the chemical potential is gauge invariant, we obtain the following relation,
\begin{eqnarray}
\mu =e\left(
\varphi + { \hbar \over {2e}}\partial_t \chi
\right)
\end{eqnarray}
With this identification, we can include the condition that the chemical potential exists in a metal.
Then, Eq.~(\ref{eq24}) is further modified as
\begin{eqnarray}
{{d{\bf v}} \over {dt}}&=&{ {-e}\over {m_e}}{\bf E}-{ {1} \over {m_e}}\nabla \mu
+{1 \over 2}\nabla v^2+{ {-e} \over m_e}{\bf v}\times{\bf B}
\nonumber
\\
&&
+{ \hbar \over {2m_e}}{\bf v}\times \nabla \times (\nabla \chi)
+{ \hbar \over {2m_e}}(\nabla \partial_t -\partial_t \nabla )\chi
\label{eq13b}
\end{eqnarray}

\section{Newton equation for the generalized Hamiltonian for electrons in electromagnetic field}
\label{sec3}

\subsection{Classical single electron case}
The classical single electron case is obtained by omitting terms with $\chi$, and also terms arising from many-body effects, $-{ {1} \over {m_e}}\nabla \mu$ and ${1 \over 2}\nabla v^2$, in Eq.~(\ref{eq13b}).
The result is 
\begin{eqnarray}
m_e{{d{\bf v}} \over {dt}}={\bf F}_{\rm Lorentz}
\end{eqnarray}
This is the formula in the usual classical electromagnetic theory.

\subsection{Many electron without multivalued $\chi$ case}

In this case, we can omit terms with $\nabla \times (\nabla \chi)$ and $(\nabla \partial_t -\partial_t \nabla )\chi$ in Eq.~(\ref{eq13b}). 
The Newton equation becomes
\begin{eqnarray}
m_e{{d{\bf v}} \over {dt}}={\bf F}_{\rm Lorentz}-\nabla \mu+\nabla {m_e\over 2}v^2
\label{classical}
\end{eqnarray}
The presence of the force $-\nabla \mu$ changes the explanation for the energy flow
occurring in a system with a metallic wire and an electrical battery, and also for that occurring in charging of a capacitor. We consider these cases, below.

\subsubsection{Steady straight current in a metallic wire connected to an electric battery.}

The battery connected to a metallic wire generates the chemical potential difference \cite{Landauer,Buttiker1985,Datta}.
In the steady straight current case, we have $\nabla v^2 =0$ in the current flow direction. The current generates ${\bf B}$ but no force acting on the conduction electrons is generated. We consider the steady flow of electrons with the compensating back ground positive charge, which yields ${\bf E} = 0$. 
We also add the term arising from resistivity using the relaxation approximation (this term is absent in Eq.~(\ref{eq13b})). Then,
the Newton equation for this case is given by
\begin{eqnarray}
m_e{{d{\bf v}} \over {dt}}=-\nabla \mu-m_e {1 \over \tau} {\bf v}=0
\label{ManyClassical}
\end{eqnarray}
where $\tau$ is the relaxation time.

In the usual explanation for the energy flow,  $-e{\bf E}$ is used instead of $-\nabla \mu$ appearing in Eq.~(\ref{ManyClassical}). Then, non-zero Poynting vector arises; it gives rise to the energy flow of radiation entering from the outside of the wire \cite{FeynmanII27-5}. This explanation is odd since  the source of the energy is chemical reactions in the battery.
The above equation explains that the energy flow occurs through the wire
due to the  generation of the chemical potential difference by the battery.
 
\subsubsection{Charging of a capacitor by an electric battery.}

The charging of a capacitor generates electric field between the electrodes of the capacitor.
In the usual explanation for the energy flow, ${\bf E}$ developed 
between the electrodes and the magnetic field generated by the displacement current arising from 
the time-dependence of ${\bf E}$ between the electrodes
generates ${\bf B}$. By these ${\bf E}$ and ${\bf B}$, non-zero
Poynting vector is produced, giving rise to the energy flow by radiation that enters from outside. Thus, 
this energy flow supplies the energy stored in the electric field of the capacitor \cite{FeynmanII27-5}.
This is odd, since the energy is supplied by chemical reactions in the battery that produce the chemical potential
difference of the two electrodes.

We can explain the energy flow in this case as follows. The vector potential in this case is calculated using the formula
\begin{eqnarray}
{\bf A}({\bf r}, t)={\mu_0 \over {4\pi}}\int d^3 r'{ {{\bf j}({\bf r}', t-|{\bf r}-{\bf r}'|/c)}
\over {|{\bf r}-{\bf r}'|}}
\end{eqnarray}
where $c$ is the speed of light; thus, only the true current ${\bf j}$ contributes to the generation of ${\bf B}$
and the displacement current generation of ${\bf B}$ is absent. 
Thus, the energy flow by the radiation does not occur. 
The electric field energy stored in the capacitor
 is supplied by the battery that generates the chemical potential
 difference between the electrodes of the capacitor.

\subsection{Many electron with multivalued $\chi$ case}

When dealing with cases where ${\bf A}$ appears explicitly as in the Meissner current in superconductors, and $\chi$ is multi-valued,
Eq.~(\ref{eq13a}) must be used. 
In this case, the Newton equation is given by
\begin{eqnarray}
m_e{{d{\bf v}} \over {dt}}={ e } \partial_t {\bf A}-{ \hbar \over {2}}\partial_t(\nabla \chi)
+{m_e \over 2}\nabla v^2-{ e }{\bf v}\times{\bf B}
+{ \hbar \over {2}}{\bf v}\times \nabla \times (\nabla \chi)
\nonumber
\\
\label{quantumForce}
\end{eqnarray}
We examine some examples that can be explained by forces  in the above equation.

\subsubsection{${ \hbar \over {2}}{\bf v}\times \nabla \times (\nabla \chi)$.}

With this term, the effective magnetic field becomes,
\begin{eqnarray}
{\bf B} \rightarrow {\bf B}-{\hbar \over {2e}}\nabla \times (\nabla \chi)
\label{breplace}
\end{eqnarray}
as infered in Eq.~(\ref{quantumForce}).
This combination of terms appears in the magnetic field equation for superconductors in the presence of vortices \cite{Abrikosov}. Using the Amp\`{e}re's circuital law 
\begin{eqnarray}
 \nabla \times {\bf B}=\mu_0 {\bf j}
 \end{eqnarray}
  where $\mu_0$ is the permeability of free space and ${\bf j}$ the current density
 \begin{eqnarray}
 {\bf j}=-e n_s {\bf v}
 \end{eqnarray}
 where $n_s$ is the superconducting electron density 
 and ${\bf v}$ the velocity field in Eq.~(\ref{eq12v2}), we have
\begin{eqnarray}
\nabla \times (\nabla \times {\bf B})= -{{e^2 n_s \mu_0} \over m_e}\left[ {\bf B} -{ \hbar \over {2e}}\nabla \times (\nabla \chi) \right]
\end{eqnarray}
Note that, the replacement in Eq.~(\ref{breplace}) appears in the right-hand-side of the equation.
This equation is simplified as
\begin{eqnarray}
\nabla^2{\bf B}= {1 \over {\lambda_L^2}}{\bf B} -{ \Phi_0 \over {\lambda_L^2}}{ 1 \over {2\pi}}\nabla \times (\nabla \chi) 
\label{abrikosov}
\end{eqnarray}
where $\Phi_0={{h} \over {2e}}$ is the flux quantum, and $
\lambda_L= \sqrt{m_e \over {\mu_0 n_s e^2}}$ is the London penetration depth.

\begin{figure}
\begin{center}
\includegraphics[width=7.0cm]{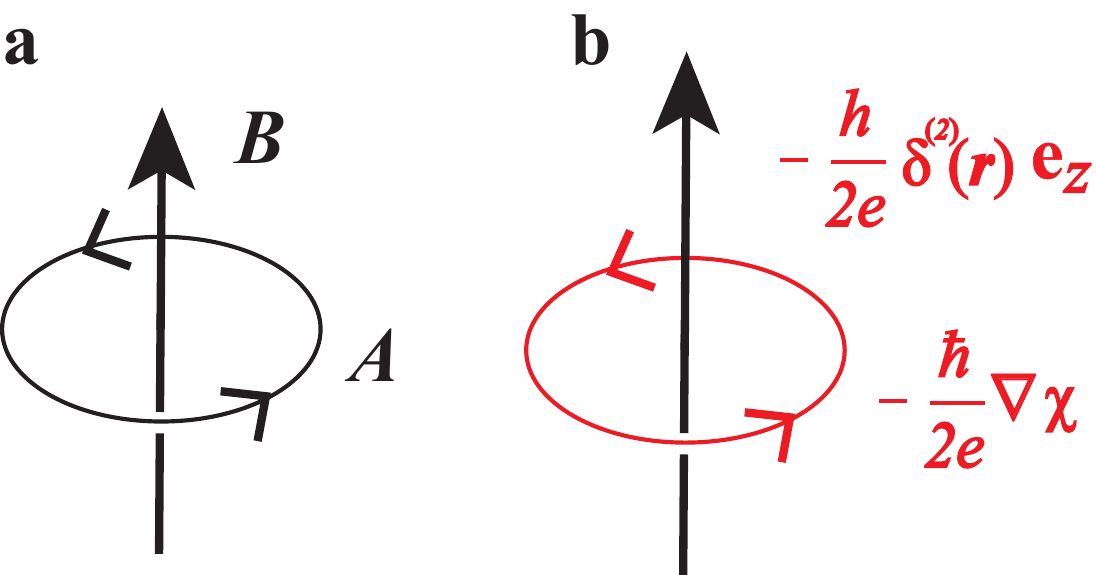}
\caption{Schematic pictures for the relation between flow and vorticity. {\bf a}: Magnetic field ${\bf B}$ is given as the vorticity of the vector potential ${\bf A}$ as ${\bf B}=\nabla \times {\bf A}$ in the Maxwell's view of the magnetic field \cite{Maxwell2}. 
{\bf b}: Vorticity arising as a delta-function like magnetic field from $\nabla \chi$. Here, the situation is considered where $\chi$ creates a winding number one vortex line centered at ${\bf r}$ along $z$-axis. There exists a string of singularities of the wave function along the vortex line. Note that such an object was considered by Dirac \cite{Monopole}.}
\label{Current-chiX}
\end{center}
\end{figure}

Maxwell interpreted that ${\bf B}$ is the vorticity of the flow ${\bf A}$ \cite{Maxwell2} (Fig.~\ref{Current-chiX}{\bf a}).
Likewise, we can consider the vorticity from the current $-{\hbar \over {2e}}\nabla \chi$.
It is expressed using a delta-function
as will be explained, below. Let us consider the case where the winding number is one, $w_C[\chi]=1$, and string of singularities exist along $z$-axis (Fig.~\ref{Current-chiX}{\bf b}).
In this case, the following relation holds
\begin{eqnarray}
{ 1 \over {2\pi}}\nabla \times (\nabla \chi) =\delta^{(2)}({\bf r}) {\bf e}_z
\label{delta}
\end{eqnarray}
where $\delta^{(2)}({\bf r})$ is a two-dimensional ($xy$-plane) delta function, and ${\bf e}_z$ is the unit vector in the $z$-direction. Thus, Eq.~(\ref{abrikosov}) becomes the formula first put forward by Abrikosov \cite{Abrikosov,TinkhamText}. It is notable that a similar string of singularities was envisaged by Dirac \cite{Monopole}. 
In the present case, those strings provide vorticity that produce loop currents.

\subsubsection{${m_e \over 2}\nabla v^2$. }

\begin{figure}[H]
\begin{center}
\includegraphics[width=8.0cm]{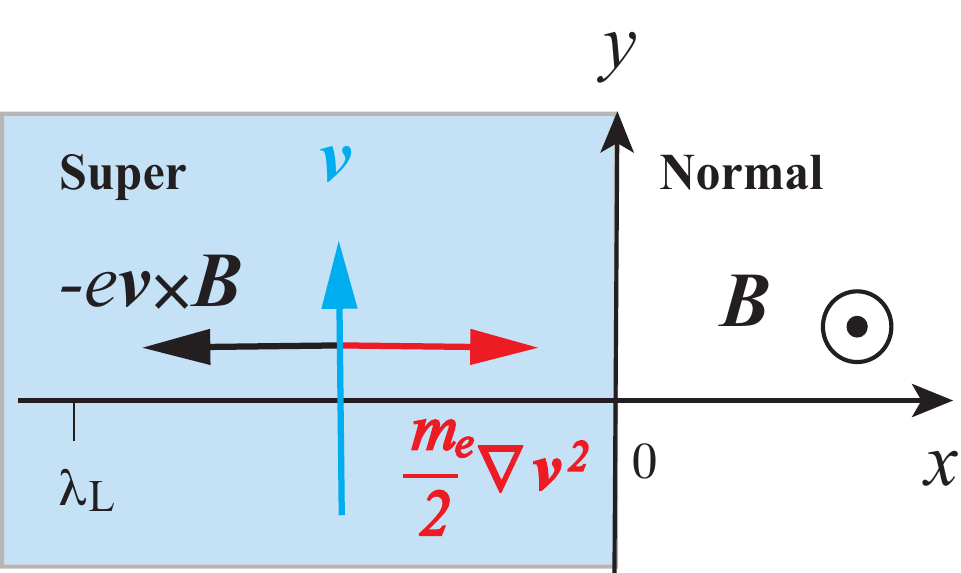}
\end{center}
\caption{A situation for the normal-superconducting phase transition in a magnetic field. The superconductor phase exists $x \le 0$, and the normal phase 
exists in $x > 0$. The magnetic field in the normal phase region is ${\bf B}=B_0 {\bf e}_z$.
The Lorentz force $-e{\bf v} \times {\bf B}$ and the gradient of the kinetic energy force ${ m_e \over 2}\nabla v^2$ balance.}
\label{S-N-inter}
\end{figure}

In a type I superconductor, a reversible superconducting-normal phase transition in a magnetic field occurs. We have discussed this before \cite{koizumi2023}, however, we reproduce it succinctly, below, for the convenience of the later discussion.
The situation we consider is depicted in Fig.~\ref{S-N-inter}. 
The magnetic field in the superconductor is given by
\begin{eqnarray}
{\bf B}=B_0 {\bf e}_z e^{ x/\lambda_L}
\label{eq11}
\label{lambda}
\end{eqnarray}
Using the Amp\`{e}re's circuital law, the current density is calculated as 
\begin{eqnarray}
{\bf j}=-{1 \over {\mu_0 \lambda_L}}B_0 {\bf e}_ ye^{ x/\lambda_L}
\end{eqnarray}
Then, the velocity field is obtained as
\begin{eqnarray}
{\bf v}={1 \over {n_s e\mu_0 \lambda_L}}B_0 {\bf e}_ ye^{ x/\lambda_L}
\label{v1}
\end{eqnarray}
From Eqs.~(\ref{eq11}) and (\ref{v1}), the following relation is obtained
\begin{eqnarray}
{m_e \over 2}\partial_xv^2=e({\bf v}\times{\bf B})_x={{B_0^2} \over {n_s \mu_0 \lambda_L}}e^{{2x} \over {\lambda_L}}
\label{eqfb}
\end{eqnarray}
This indicates that the inclusion of this term explains the force balance, correctly,
which is not possible by the standard theory that only includes the Lorentz force.

\begin{figure}[H]
\begin{center}
\includegraphics[width=7.0cm]{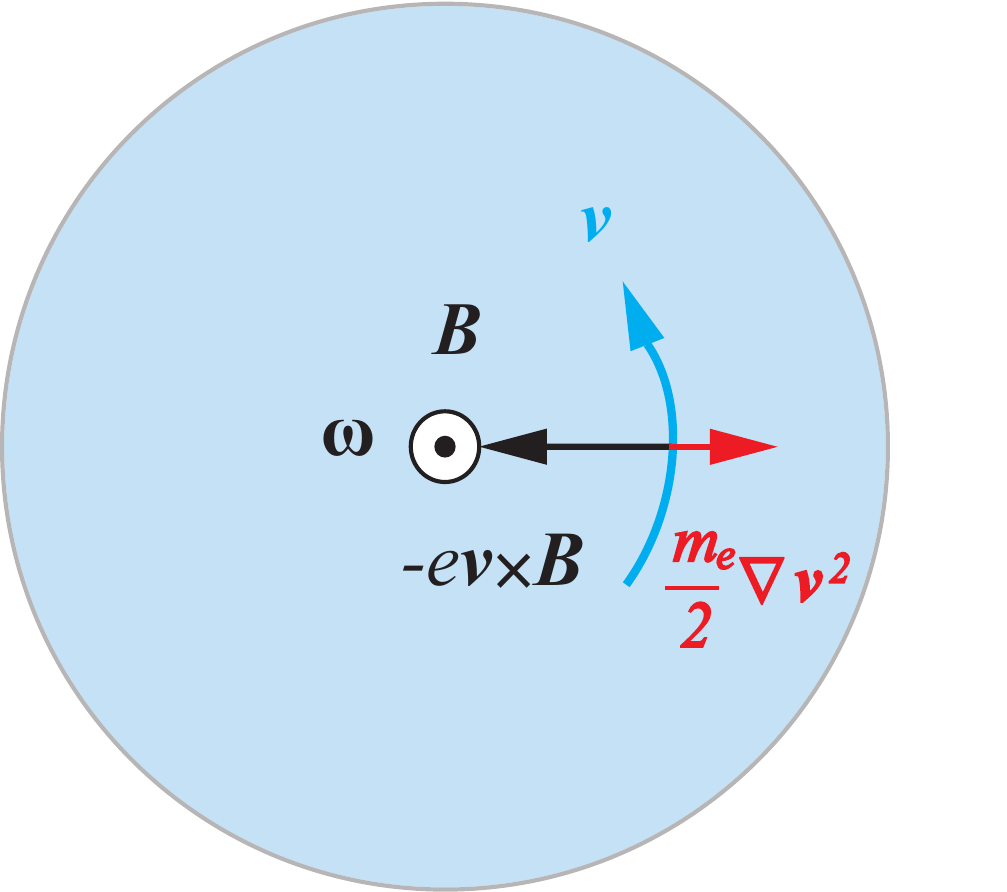}
\end{center}
\caption{The force balance of electrons for the centripetal, Lorentz, and gradient of the kinetic energy forces in the equatorial plane normal to the angular velocity ${\bm \omega }$ of a rotating spherical superconductor.
The correct centripetal force is given by $-e {\bf v} \times {\bf B}+{m_e \over 2}\nabla v^2=-{e \over 2}{\bf v} \times {\bf B}$.}
\label{LondonMoment}
\end{figure}

A similar force balance appears in the London magnetic field problem of a rotating superconductor. 
In a rotating superconductor, the velocity potential is given by 
\begin{eqnarray}
{\bf v}={\bm \omega} \times {\bf r}
\label{omega}
\end{eqnarray}
where ${\bm \omega}$ is the angular velocity of the rotation.
Then, from Eqs.~(\ref{eq12v2}) and (\ref{omega}), we have
\begin{eqnarray}
\nabla \times ({\bm \omega} \times {\bf r})=\nabla \times \left[{e \over m_e}\left({\bf A}-{\hbar \over {2e}}\nabla \chi \right)\right]
\label{omega2}
\end{eqnarray}
Neglecting delta-function type contributions that might arise from $\nabla \times (\nabla \chi)$, we have
\begin{eqnarray}
{\bf B}={{2 m_e} \over e}{\bm \omega }=-{{2 m_e} \over q}{\bm \omega }
\label{omega3}
\end{eqnarray}
where $q=-e$ is the charge of the electron. Note that the same formula is obtained for the same charge-mass ration ${m_e \over q}={{2m_e} \over {2q}}$.
This is the London magnetic field associated with the London moment phenomenon. 

The correct centripetal force for the electron cannot be obtained by the standard theory \cite{HirschLorentz,Hirsch2013b}.
However, the correct one is obtained by including the gradient of the kinetic energy term as shown below.
Let us consider a sphere of a superconductor, and consider the equatorial plane whose normal is parallel to ${\bm \omega }$ (see Fig.~\ref{LondonMoment}).
The centripetal force in this plane is in the direction toward the center, and its magnitude
is given by
\begin{eqnarray}
{{ m_e v^2} \over r }={{ m_e \omega^2 r^2} \over r }={ m_e \omega^2 r}
\label{omega4}
\end{eqnarray}
The Lorentz force $-e {\bf v} \times {\bf B}$ is in the same direction as the centripetal force, and its  magnitude is
\begin{eqnarray}
e (r \omega){{2 m_e} \over e}{\omega }={ {2m_e} \omega^2 r}
\label{omega5a}
\end{eqnarray}
The force ${m_e \over 2}\nabla v^2$ is in the opposite direction to the centripetal force; its magnitude is given by
\begin{eqnarray}
{m_e \over 2}{ d \over {dr}}(r \omega)^2={ m_e \omega^2 r}
\label{omega5b}
\end{eqnarray}
From Eqs.~(\ref{omega4})-(\ref{omega5b}), the force balance
\begin{eqnarray}
\mbox{Centripetal force}=-e {\bf v} \times {\bf B}+{m_e \over 2}\nabla v^2
\end{eqnarray}
is obtained.

\subsubsection{ $-{ \hbar \over {2}}\partial_t(\nabla \chi)$.}

Since $\chi$ is multivalued, we need to consider loop integrals of $\nabla \chi$ as given in Eq.~(\ref{Faraday4}), below. 
The multivaluedness of $\chi$ is characterized by the winding number in Eq.~(\ref{winding}). In the reversible conversion between magnetic flux and the supercurrent occurring in the reversible superconducting-normal phase transition in a type I superconductor problem, the quantum transitions changing the winding numbers will be important \cite{koizumi2024}.

The Faraday's electromotive force given in Eq.~(\ref{Faraday2}) can be expressed as the force acting on the electron fluid; namely, it can be expressed by the total time-derivative of the velocity field as
  \begin{eqnarray}
{\cal E}={1 \over {-e}}\int_{C}{{d (m_e {\bf v})} \over {dt}} \cdot d{\bf r}
\label{Faraday3}
\end{eqnarray}
This shows that  ${\cal E}$ can be regarded as the force for generating circulation
\begin{eqnarray}
\partial_t \int_{C}{\bf v}\cdot d{\bf r}
\end{eqnarray}
multiplied by $-m_e/e$.
Here, the case is assumed wheres the time-dependence of the $C(t)$ is absent.

If we consider the circulation generated by $\nabla \chi$, the term $-{ \hbar \over {2}}\partial_t(\nabla \chi)$ can be considered as the force for generating topologically protected loop currents.
Thus, we may write ${\cal E}$ as
\begin{eqnarray}
{\cal E}={1 \over {e}}\int_{C}{ {\hbar} \over 2} \partial_t (\nabla \chi) \cdot d{\bf r}
\label{Faraday4}
\end{eqnarray}
The condition for the reversible conversion between magnetic flux and the topologically protected loop current is 
given by
\begin{eqnarray}
-\oint_{C} \partial_t{\bf A}({\bf r}, t)\cdot d{\bf r}=\oint_{C} { {\hbar} \over {2e}}\partial_t \nabla \chi \cdot d{\bf r}
\label{Faraday5}
\end{eqnarray}
where the left-hand side comes from Eq.~(\ref{Faraday2}), and the right-hand side from Eq.~(\ref{Faraday4}).
The above relation may be expressed in a simplified way as
\begin{eqnarray}
\partial_t {\bf A}=-{ {\hbar} \over {2e}}\partial_t \nabla \chi
\label{JoulA}
\end{eqnarray}
This is the condition we have obtained in our previous work \cite{koizumi2024}.

\section{Concluding remarks}
\label{sec4}

 \begin{figure}
\begin{center}
\includegraphics[width=10.0cm]{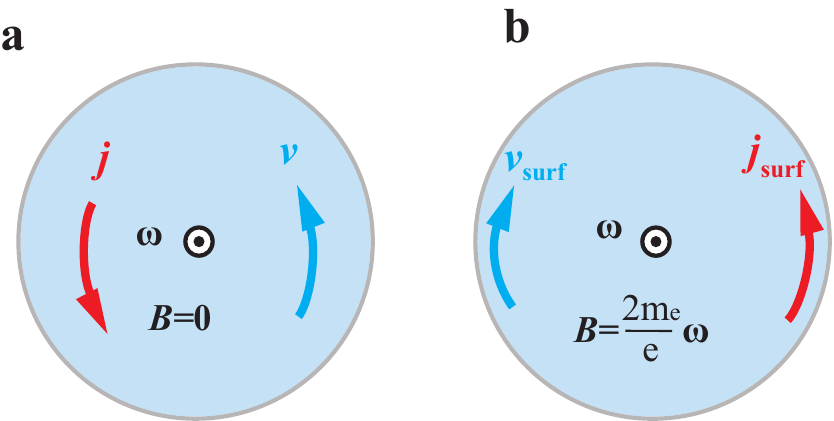}
\caption{Schematic pictures of the current in a rotating superconducting sphere that generates the London magnetic field.  {\bf a}: The velocity field for electrons ${\bf v}$ is generated by the rotation. Initially, the magnetic field is zero, ${\bf B}=0$; due to the gradient of the kinetic energy force $\nabla { m_e \over 2}{\bf v}^2$, electrons move outward direction to the surface. As a consequence the velocity of the underlying positive cores exceeds that of electrons, generating current ${\bf j}$. The generated current produces the Lorenz force that acts on electrons moving outward; eventually, the velocity of the electrons becomes that expected by the whole system rotation with ${\bm \omega}$, except in the surface region where outward motion of electrons is impossible. {\bf b}: The surface current is denoted by ${\bf j}_{\rm surf}$, which is generated due to
the slow-down of
the velocity of surface electrons by ${\bf v}_{\rm surf}$. Here,
the slow-down velocity means it is slower
than this amount from the velocity of the whole system rotation with ${\bm \omega}$.}
\label{ChiSlowdown}
\end{center}
\end{figure}

Here, we give a possible scenario how the London moment is generated from ${\bm \omega}=0$ initial state.
The explanation of this process has not been successful by the standard London theory that includes only the Lorentz force. When a spherical superconductor starts rotating from ${\bm \omega}=0$, first, the magnetic field is absent; then, electrons move outward radial direction due to the force ${{m_e} \over 2 }\nabla { v}^2 $ since the velocity of inner electrons is slower than the outer ones. This electron motion generates a magnetic field by the current creation due 
to the difference in the velocity of electrons and that of the underlying positive ion cores;
the positive ion cores move with velocity by the ${\bm \omega}$ rotation, however, that of the negative electrons moved from inner area is slower (Fig.~\ref{ChiSlowdown}). 

When the magnetic field is generated, the motion of the electrons is accelerated by the Lorentz force produced by the generated magnetic field and the outward radial direction velocity of electrons; eventually, electrons move with the rotating sphere with the same velocity as the underlying ion cores. 
This change of the electron velocity does not occur in the surface region,
since the outward radial direction motion is impossible there; therefore the electron velocity remains slower
than that of the positive ion cores; this slower electron motion gives rise to the surface current that produces the London magnetic field given by Eq.~(\ref{omega3}).

\begin{figure}
\begin{center}
\includegraphics[width=4.0cm]{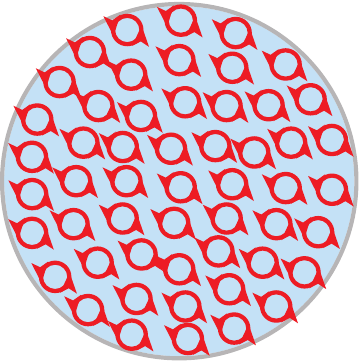}
\caption{A schematic snap shot of the velocity field of electrons that is added to the velocity from the whole system rotation in Fig.~\ref{ChiSlowdown}{\bf b}. It is assumed that it is a collection of topologically protected loop currents
produced by $\nabla \chi$, and they move with the whole system rotation; other contributions for electric current from electrons are canceled out by the electric current contribution from the background positive ion cores in a corse-grained average sence. The produced  current in the bulk is zero due to the cancelation of currents by loop currents.
Only the surface current remains.}
\label{ChiSlowdown2}
\end{center}
\end{figure}

It is also important to note that the outward radial direction motion of electrons will cause the electron density spatial variation that will generate nonzero gradient of the chemical potential. Thus, the back flow of electrons occurs.
As a consequence loop currents will be generated.
Then, the final surface current should be considered as a collection of loop currents as depicted in Fig.~\ref{ChiSlowdown2}. The bulk current is zero due to the cancelation.

The present theory indicates that the stability of $\nabla \chi$ changes the behavior of the system. If $\nabla \chi$ is not stable, the velocity in Eq.~(\ref{eq12v2}) will be zero in average, meaning
\begin{eqnarray}
\langle \oint_C {\bf A}\cdot d{\bf r} \rangle-{\hbar \over {2e}} \langle \oint_C\nabla \chi \cdot d{\bf r} \rangle=0
\end{eqnarray}
due to the fluctuation of the winding numbers for $\chi$, where $\langle  \cdots  \rangle$ denotes time-average, and $C$ is any of spacial loops in the system.
In this situation, ${\bf A}$ also fluctuates, and its explicit form is not well-defined. In this situation, ${\bf A}$ should be considered merely as a mathematical tool for the calculation.
However, if it is stable, the velocity field in Eq.~(\ref{eq12v2}) becomes meaningful with well-defined ${\bf A}$ supplemented with the compensation of the gauge ambiguity by $\nabla \chi$. In this case, persistent loop currents will flow \cite{koizumi2023,Koizumi_2023}. The electric current generated by such a velocity field should be persistent, and the Meissner effect will be realized. 
The current for this case is not the linear response to external perturbations.
Thus, it cannot be explained by the linear response theory \cite{Kubo1957}.

\appendix

\section{Definition of the velocity field operator}
\label{velocityfield}

The number density operator and the probability operator usually include Dirac delta functions.
Let us start with the number operator for an $N$-particle system; it is given by
\begin{eqnarray}
\hat{\rho}({\bf r},t)=\sum_{j=1}^N \delta({\bf r}_j-{\bf r})
\end{eqnarray}

Thus, the number density is calculated as an expectation value for $\Psi=\Psi_0 e^{ -{i \over 2} \sum_{k=1}^N \chi({\bf r}_k,t )}$ is given by
\begin{eqnarray}
{\rho}({\bf r},t)&=&
\int d{\bf x}_1 \cdots d{\bf x}_N \sum_{j=1}^N \delta({\bf r}_j-{\bf r})\Psi^\ast_0 e^{{i \over 2} \sum_{k=1}^N \chi({\bf r}_k,t)} \Psi_0 e^{ -{i \over 2} \sum_{k=1}^N \chi({\bf r}_k,t)}
\nonumber
\\
&=&N
\int d{\sigma}_1 d{\bf x}_2 \cdots d{\bf x}_N \Psi^\ast_0({\bf r}, \sigma_1, {\bf x}_2, \cdots, {\bf x}_N) \Psi^\ast_0({\bf r}, \sigma_1, {\bf x}_2, \cdots, {\bf x}_N) 
\nonumber
\\
\end{eqnarray}
Here, the fact is used that particles are identical and each contributes, equally.

The probability density current operator for an $N$-particle system is given by
\begin{eqnarray}
\hat{\bf j}({\bf r},t)=\sum_{j=1}^N {1 \over 2}\left[ \delta({\bf r}_j-{\bf r})(-i \hbar){\vec{\nabla}_j\over m_e}+(-i \hbar){{\cev{\nabla}_j} \over m_e }\delta({\bf r}_j-{\bf r})
\right]
\end{eqnarray}
Noting that the probability density current is zero for $\Psi_0$, its expectation value for $\Psi=\Psi_0 e^{ -{i \over 2} \sum_{k=1}^N \chi({\bf r}_k,t )}$ is given by
\begin{eqnarray}
{\bf j}({\bf r},t)&=&
\int d{\bf x}_1 \cdots d{\bf x}_N \sum_{j=1}^N \delta({\bf r}_j-{\bf r})\left[ -{{\hbar \nabla\chi ({\bf r}_j,t)}  \over {2 m_e}} \right]\Psi^\ast_0 e^{{i \over 2} \sum_{k=1}^N \chi({\bf r}_k,t)}\Psi_0 e^{ -{i \over 2} \sum_{k=1}^N \chi({\bf r}_k,t)}
\nonumber
\\
&=&\left[ -{{\hbar \nabla\chi ({\bf r},t)}  \over {2 m_e}} \right]N
\int d{\sigma}_1 d{\bf x}_2 \cdots d{\bf x}_N \Psi^\ast_0({\bf r}, \sigma_1, {\bf x}_2, \cdots, {\bf x}_N) \Psi^\ast_0({\bf r}, \sigma_1, {\bf x}_2, \cdots, {\bf x}_N) 
\nonumber
\\
&=&{\rho}({\bf r},t)\left[ -{{\hbar \nabla\chi ({\bf r},t)}  \over {2 m_e}} \right]
\nonumber
\end{eqnarray}

Now, the velocity operator is usually given by 
\begin{eqnarray}
\hat{\bf v}({\bf r},t)={{\hat{\bf j}({\bf r},t)} \over {\rho({\bf r},t)}}
\nonumber
\end{eqnarray}

Its expectation value is
\begin{eqnarray}
{\bf v}({\bf r},t)={{{\bf j}({\bf r},t)} \over {\rho({\bf r},t)}}=-{{\hbar \nabla\chi ({\bf r},t)}  \over {2 m_e}}
\nonumber
\end{eqnarray}

\section*{References}

\providecommand{\newblock}{}

\end{document}